\documentclass{emulateapj}
\usepackage{epsfig}
\usepackage{graphicx}

\shorttitle{Three-dimensional Doppler Images of RS Vul}
\shortauthors{Richards, Sharova, \& Agafonov}

\begin{document}

\title{Three-Dimensional Doppler Tomography of the RS Vulpeculae Interacting Binary}
\author{Mercedes T. Richards}
\affil{Department of Astronomy \& Astrophysics, Pennsylvania State University, 525 Davey Laboratory, University Park, PA, 16802, USA}
\email{mrichards@astro.psu.edu} 

\author{Olga I. Sharova}
\affil{Radiophysical Research Institute (NIRFI), 25/12a, Bolshaya Pecherskaya St., Nizhny Novgorod, 603950, Russia}
\email{shol@nirfi.sci-nnov.ru}

\and
\author{Michail I. Agafonov} 
\affil{Radiophysical Research Institute (NIRFI), 25/12a, Bolshaya Pecherskaya St., Nizhny Novgorod, 603950, Russia}
\email{agfn@nirfi.sci-nnov.ru}

\begin{abstract}
      Three-dimensional Doppler tomography has been used to study the H$\alpha$ emission sources in the RS Vulpeculae interacting binary.  The 2D tomogram of this binary suggested that most of the emission arose from the cool mass losing star with additional evidence of gas flowing close to the predicted trajectory.  However, the 3D tomogram revealed surprising evidence of a more pronounced gas stream flow at high $V_z$ velocities from -240 to -360 km s{$^{-1}$}.  This behavior is most likely caused by magnetic activity on the cool star since the central velocity plane, defined by $V_z$ = 0 km s{$^{-1}$}, should be coincident with the orbital plane of the binary if the flow is dominated by gravitational forces only.  RS Vul has been detected as both an X-ray and a radio source, and it is possible that the RS Vul gas stream may have been deflected by magnetic field lines.   This flow is distinctly different from that found in the streamlike state of U CrB, in which the gas flow was confined mostly to the central velocity plane.   
\end{abstract}


\section{Introduction}
The extension from 2D to 3D tomography is now a reality. Two-dimensional Doppler images can be created with the standard Filtered Back Projection (FBP) Method ({\it e.g.,} \citealt{kaitchucketal94}) by projecting the sources of emission onto a single central ($V_x$,$V_y$) velocity plane.  This plane is expected to coincide with the orbital plane of the binary since this is the plane in which the gravitational forces dominate the gas flows.  However, these sources may have non-zero velocities beyond the central 2D plane since they are not geometrically thin and are not confined to the orbital plane.  Instead, 3D images can be derived over a range of $V_z$ values transverse to the central plane.  The resulting 3D maps are consistent with the 2D Doppler tomograms obtained with the FBP method and reveal substantial flow structures beyond the central velocity plane.   The 3D reconstruction method applied in this work is the Radioastronomical Approach (RA) developed by \citet{agafonov04a,agafonov04b,agafonov+sharova05a,agafonov+sharova05b}.  A detailed discussion of the differences between this technique and the FBP method can be found in \citet{agafonovetal06} and \citet{agafonovetal09}.   In particular, the RA method is especially useful when the number of projections is relatively small because it can achieve the same resolution as FBP while using fewer profiles (see Figure 1 of \citealt{agafonovetal06}).   The effectiveness of the RA method has been examined by using initial 3D test models, calculating 1D profiles from these models, and then performing the reconstruction.  \citep{sharova06} compared the computed models with the initial models and showed that the computed emission maxima coincided with the initial model, and all details were reproduced.   So, there was very good agreement between the initial test model and the reconstructed 3D tomogram. 

The two-dimensional Doppler tomograms of the Algol-type binaries \citep{richardsetal95, albright+richards96,richards04} display a diverse range of circumstellar structures.  These include a gas stream, accretion annulus, transient accretion disk, shock regions and sometimes a chromospheric emission source in the short-period (P $<$ 5 days) Algols, as well as a classical accretion disk in the long-period (P $\ge$  5 -- 6 days) Algols.  \citet{richards+albright99} found that some systems, like U Coronae Borealis (U CrB) and U Sagittae (U Sge), alternate between streamlike and disklike states and display changes that can occur within a year, and sometimes overnight.  They identified RS Vul as a member of this group.   

The first applications of the 3D procedure focused on the interacting Algol-type binary U CrB.   The 2D tomograms constructed from H$\alpha$ spectra of this binary collected in 1993 and 1994 displayed two distinct patterns:  a transient accretion disk representing the disklike state (in epoch 1993) and a bright gas stream flowing along the predicted ballistic trajectory representing the streamlike state (in epoch 1994).   The 3D tomograms of these emission sources were even more distinct than the 2D versions, and both states showed evidence of a prominent emission source associated with the mass gainer, resulting from the impact of the gas stream onto the photosphere of that star \citep{agafonovetal06,agafonovetal09}.  Chromospheric emission from the mass loser was also visible especially when the binary was in the disklike state.    The 3D image also showed distinct evidence of gas flows (jets) with high $V_z$ velocities.  These jets were associated with three emission sources: (1) the impact/splash region where the gas stream strikes the mass gainer in the streamlike state ($V_z$ $\sim$ -300 to +100 km~s$^{-1}$); (2) the interaction between the gas stream and accretion disk in the disklike state ($V_z$ $\sim$ 300 to 400 km~s$^{-1}$); and (3) the Localized Region where the disk material interacts with the incoming gas stream after traveling around the mass gainer ($V_z$ $\sim$ +200 to +500 km~s$^{-1}$).  These high values are expected since the star-stream impact occurs at speeds of up $\sim$500-600 km~s$^{-1}$, compared to the slower rotation of the mass gainer \citep{richards92, blondinetal95, richards+ratliff98}.  Moreover, the pattern of $V_z$ velocities associated with the disk emission implies that the disk may be inclined to the orbital plane; which subsequently suggests that the rotation axis of the mass gainer could be precessing as a consequence of the star-stream impact.    This result would not have been possible from 2D tomograms alone.

In this paper, we have continued our exploration of 3D tomography by studying the RS Vul system.   Specifically, we have 1) identified the features in the 3D images and related them to those seen in the 2D tomograms; 2) compared the 3D tomogram of RS Vul with those of U CrB; and 3) provided a physical interpretation of these features.   The system parameters and 2D results are described in Section 2, the 3D tomograms are described in Section 3, a model for RS Vul is given in Section 4, and the Conclusions are outlined in Section 5.

%Section 2
\section{2D Doppler Tomography of RS Vul} 

The Algol-type binary RS Vulpeculae is an interacting system in which the cooler G1 III secondary star is transferring mass to its B5 main-sequence companion (the primary) through Roche lobe overflow.  RS Vul is a short-period Algol ($P_{orb}$ = 4.4776635 days; \citealt{holmgren89}) with an orbital inclination of 78.7$^\circ$ \citep{hutchings+hill71}.    The properties of the system are $M_p$ = 6.59 $M_\odot$, $M_s$ = 1.76 $M_\odot$, $R_p$ = 4.71 $R_\odot$, $R_s$ = 5.84 $R_\odot$, systemic velocity $V_o$= -20.1 km s{$^{-1}$}, velocity semi-amplitude $K_p$ = 54.0 km s{$^{-1}$}, with a mass ratio, $q$=0.27 \citep{holmgren89}.  Eighty-one H$\alpha$ (6562.8 \AA) spectra with a resolution of 0.166 \AA/pixel  or 7.6 km s{$^{-1}$} were collected in 1993 with the KPNO 0.9m Coud{\'e} Feed Telescope (see \citealt{richards+albright99}). The spectra were collected at closely spaced positions around the entire orbit of the binary.  Since the observed spectra are dominated by the spectrum of the B5 mass gaining star, model atmospheres calculations were used to represent the photospheric contributions of the stars, and these stellar contributions were subtracted from the observed spectra to create difference spectra (see \citealt{richards+albright99} for details).   The difference spectra were used to calculate the tomograms.

The standard 2D Doppler tomogram of RS Vul 1993 is shown in Figure 1 along with the 2D tomogram of U CrB in the streamlike state (from \citet{richards01}).   These 2D images were compared with the 3D Doppler tomograms based on the same data.  In the tomograms, the solid trajectory is the gravitational free-fall path of the gas stream; and the circles along this trajectory are marked at intervals of a tenth of the distance from the L$_1$ point to the distance of closest approach to the mass gainer.  The largest solid circle and the smaller dashed circle mark the inner and outer edge of a Keplerian disk, respectively; the asterisk is the predicted location where the gas stream should strike the photosphere of the mass gainer; and the plus sign marks the center of mass of the binary.   The RS Vul tomogram shows the presence of the gas stream, a circumprimary emission source, chromospheric emission, and a region dominated by absorption called the absorption zone \citep{richards01}.  The strongest emission source corresponds to the chromospheric H$\alpha$ emission \citep{richards+albright96}.

%Section 3
\section{3D Doppler Tomography}

%Section 3.1
\subsection{Reconstruction of the 3D Tomogram}

For the reconstruction of the 3D Doppler tomograms, we analyzed the same set of H$\alpha$ line profiles that were used by \citet{richards01} to construct the standard 2D Doppler tomogram of RS Vul.  The 3D Doppler tomogram was calculated as a set of numerical values in the cells of a cube, with dimensions $(V_x,V_y,V_z)$ having values ranging from -700 to +700 km~s$^{-1}$.  The maximum of this function was normalized to the unit to permit the visualization of the image based on any slice taken from any direction.  The velocity resolution of the image depends on the number of projections (i.e., number of spectra) and their distribution in orbital phase; and the orbital inclination influences the ratio of the resolutions in the $V_x$, $V_y$, and $V_z$ directions.  The wavelength resolution of the spectra is 7.6 km~s$^{-1}$, so it plays a smaller role in setting the resolution of the image since it is already sufficient to produce the best possible velocity resolution for the tomogram.  In particular, the velocity resolutions in the ($V_x$,$V_y$), ($V_z$,$V_x$), or ($V_z$,$V_y$) planes are equal only for an inclination of 45$^\circ$; and the resolutions in the $V_x$ and $V_y$ directions are better than that in the $V_z$ direction for high values of inclination.  Since the orbital inclination of RS Vul is high ($i$ = 78.7$^\circ$), the 3D tomogram was restored with a resolution of 30 km~s$^{-1}$ in the $V_x$ and $V_y$ directions and 105 km~s$^{-1}$ in $V_z$ direction.  Since the resolution in the $V_z$ direction is approximately 3.5 times lower than in the other directions, features on the tomogram may look elongated in this direction.  The appearance of the images has been simplified by displaying the ($V_x$,$V_y$) slices in the horizontal plane to be consistent with the direction of the  orbital plane of the binary, and the slices containing the $V_z$ axis are in the vertical direction. 

A direct comparison was made between the observed spectra and those computed from the reconstructed 3D Doppler tomograms. Figure 2 shows this match for the 3D Doppler tomograms of RS Vul. The radial velocity, $V_r$ is plotted versus orbital phase, with a resolution of 30 km~s$^{-1}$.  The observed spectra are displayed in the left frames and the spectra computed from the 3D Doppler tomograms are displayed in the middle frames.    Two S-wave patterns are noticeable in the observed and computed spectra corresponding to the motions of the two stars.  The overall agreement between the observed and computed spectra is very good.   The quality of the fit is shown in the right frame of Figure 2, which displays the relative chi-square statistic, $\chi^2/\chi_c^2$, versus orbital phase.  Here, $\chi^2$ is normalized to the critical value of $\chi_c^2$, which corresponds to the largest acceptable value of $\chi^2$ at the 99\% confidence level (e.g., \citealt{skilling+bryan84}). Figure 2 shows that $\chi^2$ was less than 20\% of the critical value at most orbital phases and it was only as high as 65\% of the critical value over a narrow phase range from 0.47 to 0.49, perhaps because the 3D code assumes that the gas is optically thin when it might be optically thick at these phases.  So, in all cases, $\chi^2$ was much lower than the critical value which confirms that the 3D reconstruction has produced a reliable computed model for RS Vul.

%Section 3.2
\subsection{Description and Interpretation of Emission Features} 

RS Vul and U CrB have similar geometries, so we expect to identify similar features in their Doppler images.  Figure 3 displays 20 slices in the ($V_x$,$V_y$) plane for values of $V_z$ ranging from $-540$ km~s$^{-1}$ to $+540$ km~s$^{-1}$, in 60 km~s$^{-1}$ steps.  The superposition of the color or gray-scale images with the contour images allows us to emphasize the main features in each $V_z$ slice of the 3D image.   Figure 4 displays the images of seven representative $V_z$ slices over the range from $V_z$ = $-480$ km~s$^{-1}$ to $+480$ km~s$^{-1}$.  A similar figure for U CrB in the streamlike state is shown in Figure 5 for comparison.  Figure 6 displays several cross-sections of the flow in the ($V_y$,$V_z$) plane.  In Figures 3 -- 6, the light (yellow) areas with blue contours represent absorption while the dark (deep pink) areas with red contours represent emission.  

\citet{agafonovetal09} identified several features in the 3D tomogram of U CrB that were consistent with those found in the 2D image studied by \citet{richards01}:
(1) {\it circumprimary emission} (centered on the mass gainer) and an {\it accretion annulus} (a circular region of low velocity);
(2) {\it chromospheric emission} from the cool donor star (the secondary); 
(3) the {\it gas stream} flowing along its predicted ballistic trajectory;  
(4) the {\it star-stream impact region}, where the gas stream strikes the stellar surface; 
(5) emission within the {\it predicted locus of the accretion disk};
(6) a {\it localized region} between the stars where the gas stream strikes material that has circled the mass gainer;
(7) a second {\it localized region} where the gas stream makes impact with the outer edge of the accretion disk; 
(8) a {\it jet or flare} associated with the chromosphere of the secondary star; and 
(9) a {\it high-velocity stream or jet} extending in the Vz-direction beyond the stream-star impact region.   

The 3D image of RS Vul displays all of the emission sources found in U CrB (see Figs. 3 and 4), except that there was almost no emission from an accretion disk and little evidence of the star-stream impact region.   Figure 3 shows that the strongest emission sources are associated with the cool mass loser as well as the mass gainer.  The emission sources are described in Table 1 along with their typical velocity ranges and a brief description of each feature is provided below. 

\noindent
1.  The {\it circumprimary emission} centered on the velocity of the mass gainer in RS Vul is labeled ({\bf 1}) in Figure 4.   It is the strongest source of emission in the 3D image (see Figure 3, slice at $V_z$ = $-160$ km~s$^{-1}$).  This feature can be seen in the slices with $V_z$ velocities from 0 to -320 km~s$^{-1}$ in Figures 3 and 6 (see also Table 1).  Moreover, it is distinctly different from the same feature seen in U CrB 1993, which displayed a symmetric $V_z$ distribution from $V_z$ = $-180$ to $+180$ km~s$^{-1}$ (see Table 1 in \citealt{agafonovetal09}).  The circumprimary emission has higher velocities than expected from synchronous rotation, so the star may have been spun up by the impact of the gas from the donor star.

\noindent
2. {\it Chromospheric emission} from the cool mass losing star is feature ({\bf 2}) and it is comparable in strength to the circumprimary emission.  It is associated with a wide range of $V_z$ velocities from $-530$ to $+350$ km~s$^{-1}$ (see Figures 3, 4, 6, and Table 1).  This range is illustrated in Figure 6 and resembles a jet-like feature in the $V_z$ direction.   The feature has the same velocity as the mass losing star.

\noindent
3. The {\it  gas stream} flowing along its predicted ballistic trajectory is feature ({\bf 3}).  It flows from the L1 point towards the mass gaining star with speeds up to 450 km~s$^{-1}$ in the ($V_x$,$V_y$) plane, with a $V_z$ velocity of $-300$ km~s$^{-1}$.  RS Vul is unusual because the flow along the gas stream is not strongest in the $V_z$ = 0 km~s$^{-1}$ frame but in the $V_z$ = -240 to -360 km~s$^{-1}$ range (see Figs. 3 and 4).   This is distinctly different from U CrB in its streamlike state (Fig. 5), in which the gas stream flow was strongest in the central velocity plane ($V_z$ = 0 km~s$^{-1}$) and located directly on the predicted trajectory of the gas stream in the ($V_x$,$V_y$) plane.  This is not the case for RS Vul.

\noindent
4. The predicted location of the {\it star-stream impact region}, where the gas stream strikes the stellar surface, is found at ({\bf 4}) near the location where the gas stream trajectory intersects the inner part of the disk.   Even in its most extended state (when it is at $V_z$ = -300 km~s$^{-1}$), the gas stream (Feature {\bf 3}) never gets to this predicted velocity, but is truncated at the velocity corresponding to that of Keplerian flow at the surface of the mass gainer. 

\noindent
5. There is very little evidence of any emission within the {\it predicted locus of the accretion disk} (Feature {\bf 5}) in the RS Vul images, although some clumping is seen within this region.  This feature was much stronger in U CrB, and there is slightly more space around the mass gainer in U CrB to form a disk.  However, gas may have traveled around the mass gainer and created the localized regions (described below) when the interaction with the stream slowed the gas flow.   

\noindent
6. A {\it Localized region (LR)} between the stars labeled ({\bf 6}) was found near the center of the tomogram where the flow around the mass gainer is expected to strike the inner edge of incoming gas stream (see Cartesian models of $\beta$ Per \citep{richards92,richards93}, RW Tau \citep{vesper+honeycutt93}, and other direct-impact systems).   In RS Vul, this region is detected at high positive $V_z$ velocities ranging from $+50$ to $+300$ km~s$^{-1}$ and seems to develop into the structure seen at the L1 point from  $V_z$ = $+360$ to 540 km~s$^{-1}$ (Fig. 3).  The classical localized region between the stars is seen along the $V_x$=0 km~s$^{-1}$ line for $V_z$ velocities from 50 to 540  km~s$^{-1}$ in Figure 3.  However, it is difficult to say whether the part with L1 point velocities in the ($V_x$,$V_y$) plane for $V_z$ = $+180$ to $+300$ km~s$^{-1}$ (Fig. 4) should be considered part of the cool star or part of the LR.    This feature is also seen in Figure 6 slices for  $V_x$ = $-20$ to $+20$ km~s$^{-1}$.  
This kind of braking was first suggested by Richards (1992) in the study of  the Algol-type binary $\beta$ Persei and confirmed using hydrodynamic simulations of \citet{blondinetal95} and \citet{richards+ratliff98}.  Similar results were found later from gas dynamical modeling of cataclysmic variables by \citet{kuznetsovetal01}, \citet{bisikaloetal00a}, and \citet{bisikaloetal00b}. 

\noindent
7.  A second compact region labeled ({\bf 7}) also has high $V_z$ velocities from 0 to 350 km~s$^{-1}$ and is located on the $V_y$ = 0 km~s$^{-1}$ line (see Figures 3 and 4).  In Figure 6, it can be seen in the $V_x$ = 100 to 120 km~s$^{-1}$ frames, with an average $V_z$ velocity of about 250 km~s$^{-1}$.   This feature may later merge with the other localized region that runs along the line of centers between the stars.

\noindent
8. The feature labeled ({\bf 8}) has a velocity similar to the $V_x$ and $V_y$ velocities of the donor star, but with high positive $V_z$ velocities in the range from $V_z$=$+180$ to $+540$ km~s$^{-1}$ (see Figures 3 and 4).  It has $V_x$ and $V_y$ velocities of 100 to 300 km~s$^{-1}$ (see Figs. 3, 4, and 6).   This feature could be the result of chromospheric emission on the cool mass loser, and the high velocities suggest that it could correspond to a jet or flare.  However, another explanation is that it is related to the interaction between the inner edge of the gas stream and the material that has circled the mass gainer.  A similar emission source was found in U CrB.

\noindent
9. The absorption feature in RS Vul seen in the lower left and right quadrants of Figure 3 from $V_z$ = -120 to +540 km~s$^{-1}$ was called the {\it absorption zone} by \citet{richards01,richards04}, and is associated with the locus where the gas temperature becomes too high to emit at H$\alpha$.  A structure at a similar location was identified in the ultraviolet tomogram of U Sge by \citet{kempner+richards99} and in the H$\alpha$ tomogram of U CrB by \citet{agafonovetal09}. 

The 3D tomograms also reveal an interesting symmetry in the emission sources found in the $V_z$ slices.   A strong source near the mass losing star (in velocity space) is detected with intensities of 0.852 and 0.955 in symmetrical locations at $V_z$ = $-300$ and $+300$ km~s$^{-1}$, respectively (see Figure 4).  A similar symmetric intensity pattern (I = 0.343 and 0.333) is seen at $V_z$ = $-480$ and $+480$ km~s$^{-1}$.   In the case of the mass gaining star, another symmetric pattern is revealed.  The most intense emission feature in the $V_z$ = 180 km~s$^{-1}$ frame in Figure 4, with I=0.612, is located along the $V_x$ = 0 km~s$^{-1}$ line at $V_y$ = $+80$ km~s$^{-1}$, in the opposite $V_y$ direction from the circumprimary emission.  These symmetries may be purely coincidental or perhaps we can see the same feature with motions on both sides relative to the central velocity plane, with a $-V_z$ component in $-z$ direction and with a $+V_z$ component in $+z$ direction.   Nevertheless, it is expected that the 
gas flows should be symmetric about the central velocity plane if gravitational forces dominate the gas flows.  So, the displacements of emission features {\bf (1)}, {\bf (2)}, and {\bf (3)} from the central velocity plane should not be unusual.   

%Section 4
\section{A Model for RS Vul}

Figure 7 displays the 3D tomograms of RS Vul and U CrB in a 3D velocity cube together with the 2D integral images for the main planes: ($V_x$,$V_y$), ($V_x$,$V_z$), and ($V_y$,$V_z$).  These 2D integral images represent the integrals along the  $V_z$, $V_y$, and $V_x$ directions, respectively; and   the image in the ($V_x$,$V_y$) plane is equivalent to the 2D Doppler tomogram.   The U CrB images in this figure are derived from the combined disk and stream states of that system.   The 3D velocity cube of RS Vul illustrates that RS Vul has no gas stream emission along the predicted trajectory in its central velocity plane, although the 2D integral image in the ($V_x$,$V_y$) plane shows that the gas stream emission is indeed prominent, but at $V_z$ velocities much higher than expected.   In addition, both binaries display a gas velocity distribution that is inclined to the vertical direction ($V_x$ = 0 km~s$^{-1}$) in the 2D integral ($V_x$,$V_z$) plane, but vertical in the 2D integral ($V_y$,$V_z$) plane. 

A pattern has emerged from the study of the 3D tomogram of RS Vul.   Both the circumprimary emission source labeled earlier as {\bf (1)} and the gas stream {\bf (3)} were offset toward negative $V_z$ velocities, while the chromospheric emission {\bf (2)} had $V_z$ velocities over the range from $-530$ to $+350$ km~s$^{-1}$ (see also Figure 6).   So, the circumprimary emission was strong over the same range of $V_z$ velocities corresponding to the strongest gas stream emission.   Moreover, the gas stream emission extended to the Keplerian velocity at the photosphere of the mass gainer, while the mass gainer was spinning faster than the synchronous value.   Therefore, the results are consistent with the assumption that the circumprimary emission (from a photospheric bulge) is produced when the gas stream strikes the surface of the star.   In addition, the mostly negative $V_z$ velocities for the gas stream and the circumprimary emission suggest that the gas does not move in the central velocity plane, as expected if gravitational forces dominate the gas flows, and so the flow around the mass gainer is concentrated on one side of the central velocity plane.

While the image reconstruction process can now produce 3D velocity images, we are still unable to create the desired 2D or 3D Cartesian images because the various emission features have different velocity fields.   However, an estimate of the gas stream flow can be made from information provided in Figure 1. Based on the point where the gas stream is truncated, we can produce a Cartesian model to show the dominant emission sources in  RS Vul.  The distance from the L$_1$ point to the center of the primary star has been divided into 10 equal segments in the center frame of Figure 1, and the similar small circles on the Doppler tomograms are marked at velocity intervals corresponding to those distance intervals.   Comparing the center and right frames of Figure 1, we see that the gas stream flow is truncated after traveling 7 small circles on the gas stream trajectory, which is equivalent to 70\% of the distance from the L$_1$ point to the center of the primary star.  The star-stream interaction occurs along the path of the gas stream right where the gas stream ends (at 70\% of the distance from the L$_1$ point).  Both Figures 3 and 4 show that the gas flow reaches  the stellar surface corresponding to the solid circle in the tomogram, which corresponds to the Keplerian velocity at the surface of the mass gainer.  Beyond this point, there is no evidence of any significant disk in this binary at this epoch. 

Evidence of circumstellar gas distributed well above and below the orbital plane of the binary has been derived from studies of ultraviolet resonance lines and the H$\alpha$ line of several Algol systems ({\it e.g.,} Peters and Polidan 1984, Richards 1993).  This gas can reach heights comparable to the radius of the mass gaining star.  However, semi-analytical ballistic studies ({\it e.g.,} Lubow and Shu 1975, 1976) or hydrodynamic simulations of the mass transfer process ({\it e.g.,} \citealt{richards+ratliff98}) assume that the gas should be found very close to the orbital plane, if magnetic fields are not involved.    However, \citet{sternetal92} suggested that coronal mass ejections could account for more than 10{\%} of the mass transfer expected solely from Roche lobe overflow by gravitational processes alone.  The coronal behavior of the cool stars in Algol binaries is enhanced because these stars rotate more rapidly than the slowly rotating Sun.  If we can scale the solar coronal behavior to the enhanced levels seen in many Algol binaries, then radio and X--ray flares detected from these systems can be interpreted as the periodic reconnection of the coronal field \citep{sternetal92}.  RS Vul was detected as an X-ray source by \citet{white+marshall83} with an X-ray luminosity, $L_x = 2.0 \times 10^{30}$ erg s$^{-1}$, which is about half of the x-ray luminosity of $\beta$ Per.  Radio emission from RS Vul was also detected by \citet{umanaetal98} with a flux density of 0.26 mJy and radio luminosity, $L_{radio} = 3.7 \times 10^{16}$ ergs cm$^{-2}$ s$^{-1}$ Hz$^{-1}$, compared to 0.30 mJy and $9.1 \times 10^{16}$ ergs cm$^{-2}$ s$^{-1}$ Hz$^{-1}$ for U CrB.  So, the magnetic fields of the active stars in RS Vul and U CrB should be comparable.  

The long-term radio flare survey of the prototype Algol system $\beta$ Per by \citet{richardsetal03} demonstrated that major flares (up to 1Jy)  occur regularly and predictably every 48.9 $\pm$ 1.7 days (or 17 orbital cycles), thus the flaring rate is at least 7.5 yr$^{-1}$ and could be as high as 100 yr$^{-1}$ (for weaker flares every orbit).  Mass ejections from X-ray flares observed with EXOSAT and GINGA are estimated to be $\simeq$ 5 -- 50 $\times 10^{-15}$ $M_\odot$ \citep{sternetal92}, and a total mass ejection rate of $\simeq$ 3.75 -- 37.5 $\times 10^{-14}$ $M_\odot$~yr$^{-1}$ would correspond to the detected flaring rate of 7.5 yr$^{-1}$ for $\beta$ Per.  These arguments suggest that coronal mass ejections may influence the  transfer of gas between stars and even the angle at which the flow is directed.  The differences between the gas distributions in RS Vul and U CrB cannot be readily explained by differences in magnetic activity based on their x-ray and radio luminosities.  However, it is evident that magnetic fields may play a role in the distribution of circumstellar gas in both systems.

%4. CONCLUSIONS
\section{Conclusions}

Three-dimensional Doppler tomography has been used to study the H$\alpha$ emission sources in the RS Vulpeculae interacting binary and has led once again to the discovery of gas flows with significant $V_z$ velocities.   The 2D tomogram of this binary suggested that most of the emission arose from the cool mass losing star with additional evidence of gas flowing close to the predicted trajectory.  However, the 3D tomogram revealed surprising evidence that a more pronounced flow of gas along the stream was not found as expected in the central velocity plane ($V_z$ = 0 km s{$^{-1}$}) but at $V_z$ velocities from -240 to -360 km s{$^{-1}$}.  Magnetic activity on the cool mass losing star may explain the deflection of the gas stream flow away from the orbital plane of the binary where the gravitational forces dominate the interactions between the stars.   This flow behavior is distinctly different from that found in the streamlike state of U CrB, in which the gas flow was confined mostly to the central velocity plane.  Exploration of the 3D velocity structures of other binary systems should improve our understanding of these out-of-plane gas motions.

\acknowledgements
We thank the referee for comments on the manuscript.   This research was partially supported by the Russian Foundation for Basic Research (RFBR) grants 06-02-16234 and 09-02-00993, and National Science Foundation grant AST-0908440.

\clearpage

\begin{figure*}
\figurenum{1}
\center
\epsscale{0.9}
\includegraphics[height=60mm]{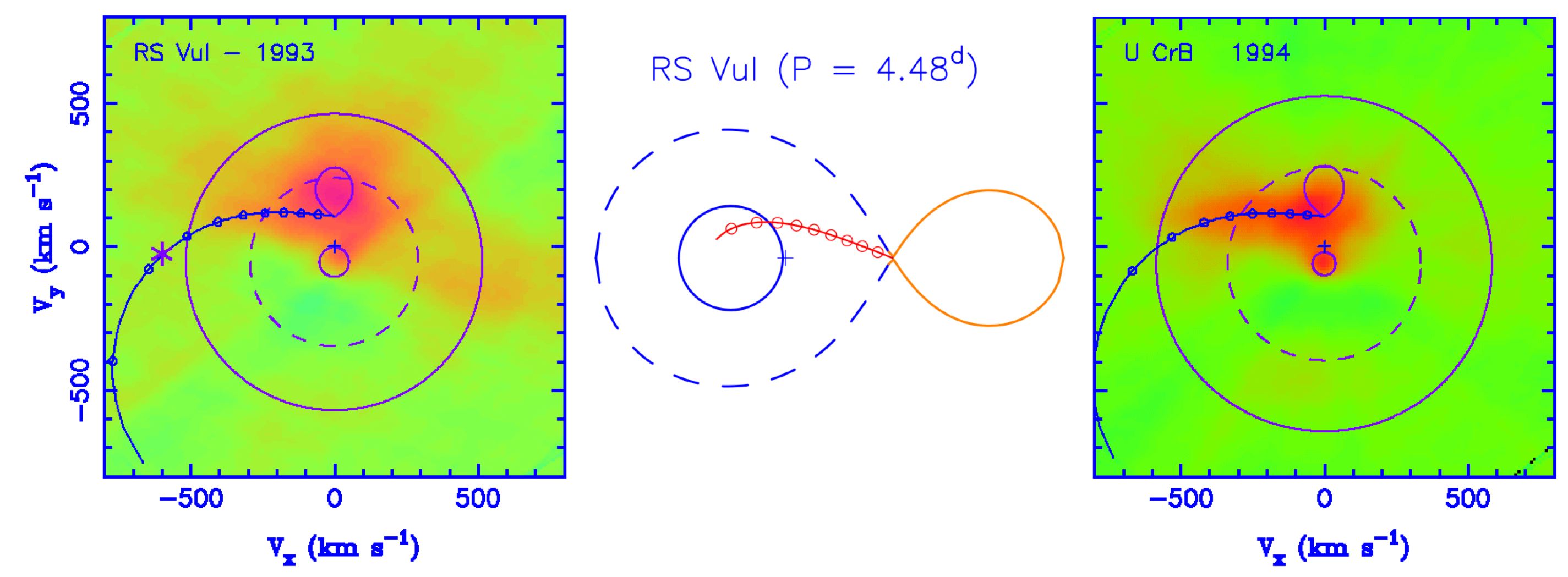}
\caption{Two-dimensional Doppler tomogram of RS Vul - 1993 (left) taken from \citet{richards01}, a Cartesian model of RS Vul (center), and the 2D tomogram of U CrB in the streamlike state (right). In the tomograms, the solid trajectory is the gravitational free-fall path of the gas stream; and the circles along this trajectory are marked at intervals of a tenth of the distance from the L$_1$ point to the distance of closest approach to the mass gainer.  The largest solid circle and the smaller dashed circle mark the inner and outer edge of a Keplerian disk, respectively; the asterisk is the predicted location where the gas stream should strike the photosphere of the mass gainer; and the plus sign marks the center of mass of the binary.
}
\label{f1}
\end{figure*} 

\begin{figure*}
\figurenum{2}
\center
\epsscale{0.9}
\includegraphics[height=75mm]{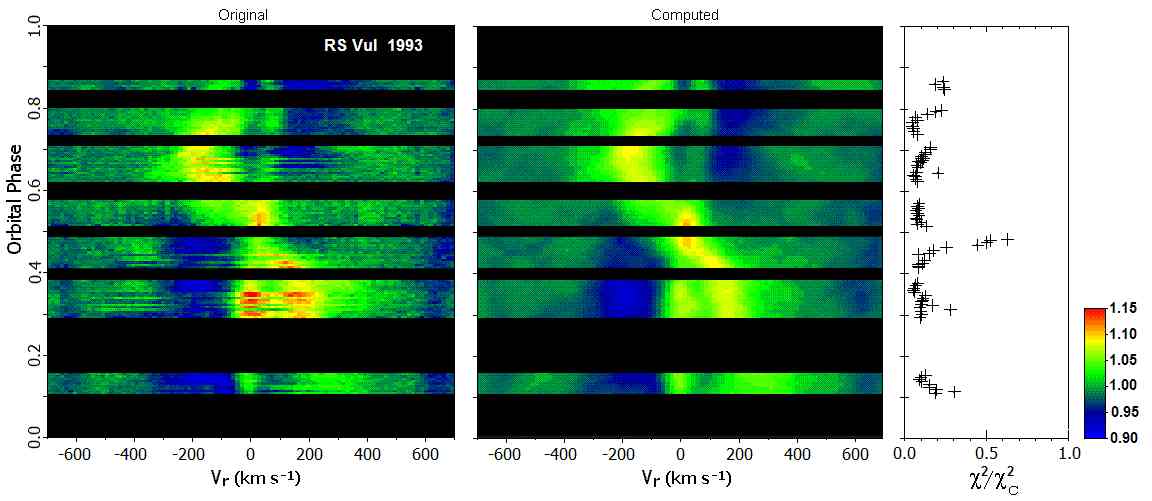}
\caption{Comparison between the original data (left frame) and the spectra computed from the reconstructed 3D Doppler map of RS Vul (middle frame) in terms of the radial velocity, $V_r$ versus orbital phase.  The right frame displays the orbital phase variation of the relative chi-square statistic, $\chi^2/\chi_c^2$,  where $\chi_c^2$ is the critical value corresponding to the 99\% confidence level.  The agreement between the observed and computed spectra is very good.
}
\label{f2}
\end{figure*}

\begin{figure*}
\figurenum{3}
\center
\epsscale{1}
\includegraphics[height=200mm]{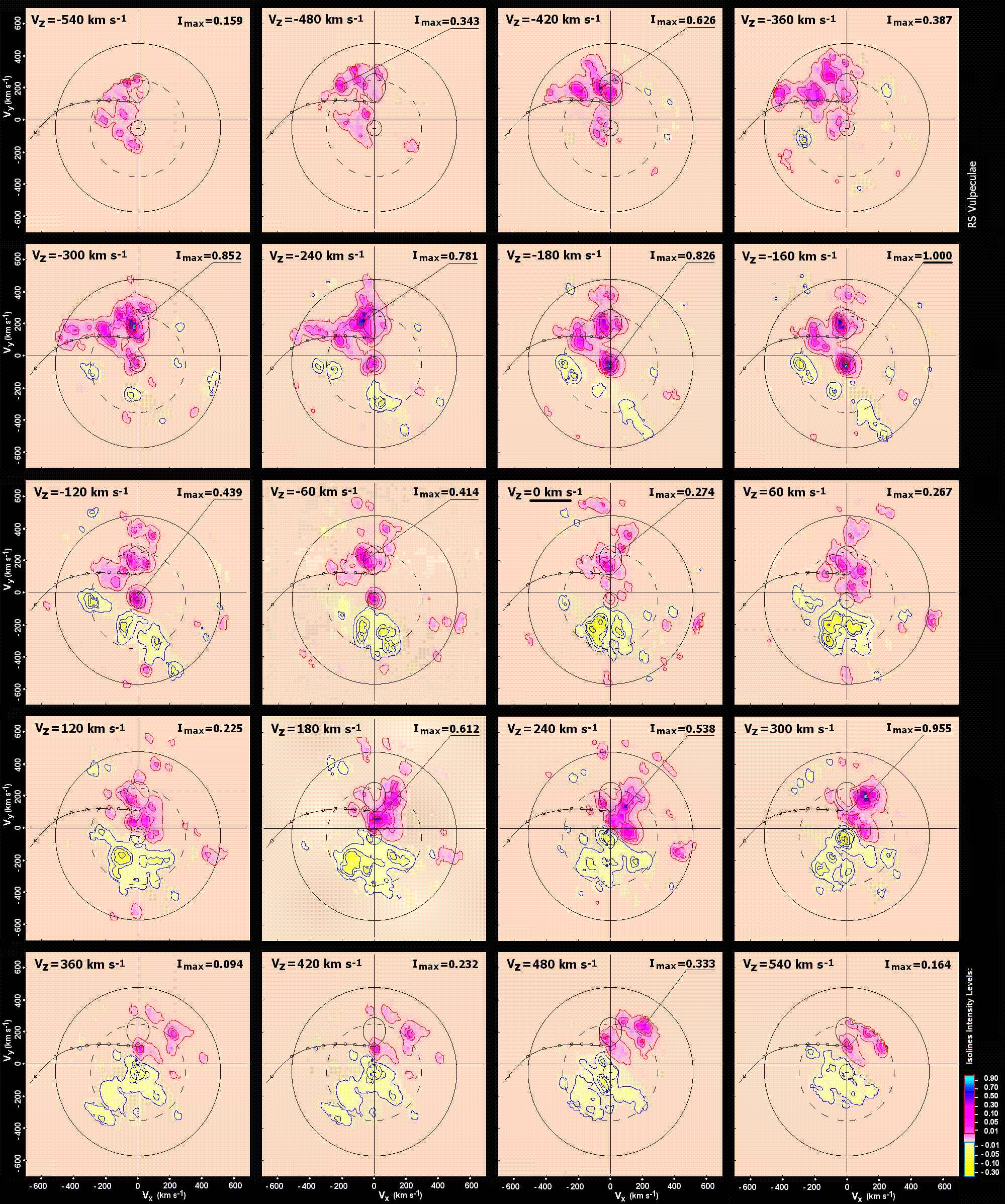}
\caption{($V_x$,$V_y$) slices of the 3D Doppler tomogram of RS Vul, for velocities beyond the central velocity plane from $V_z$ = -540 km~s$^{-1}$  to $V_z$= +540 km~s$^{-1}$, in steps of 60 km~s$^{-1}$. The images are shown for intensity levels: +0.01, 0.05, 0.1, 0.3, 0.5, 0.7, 0.9 (red) and -0.01, -0.05, -0.1, -0.3 (blue). The central source, corresponding to the mass gainer, is strongest at $V_z$= - 160 km~s$^{-1}$  and very weak at Vz= 0 km~s$^{-1}$, while the gas stream emission is strongest from $V_z$= -360 km~s$^{-1}$  to $V_z$= -240 km~s$^{-1}$ .
}
\label{f3}
\end{figure*} 

\begin{figure*}
\figurenum{4}
\center
\epsscale{0.7}
\includegraphics[height=200mm]{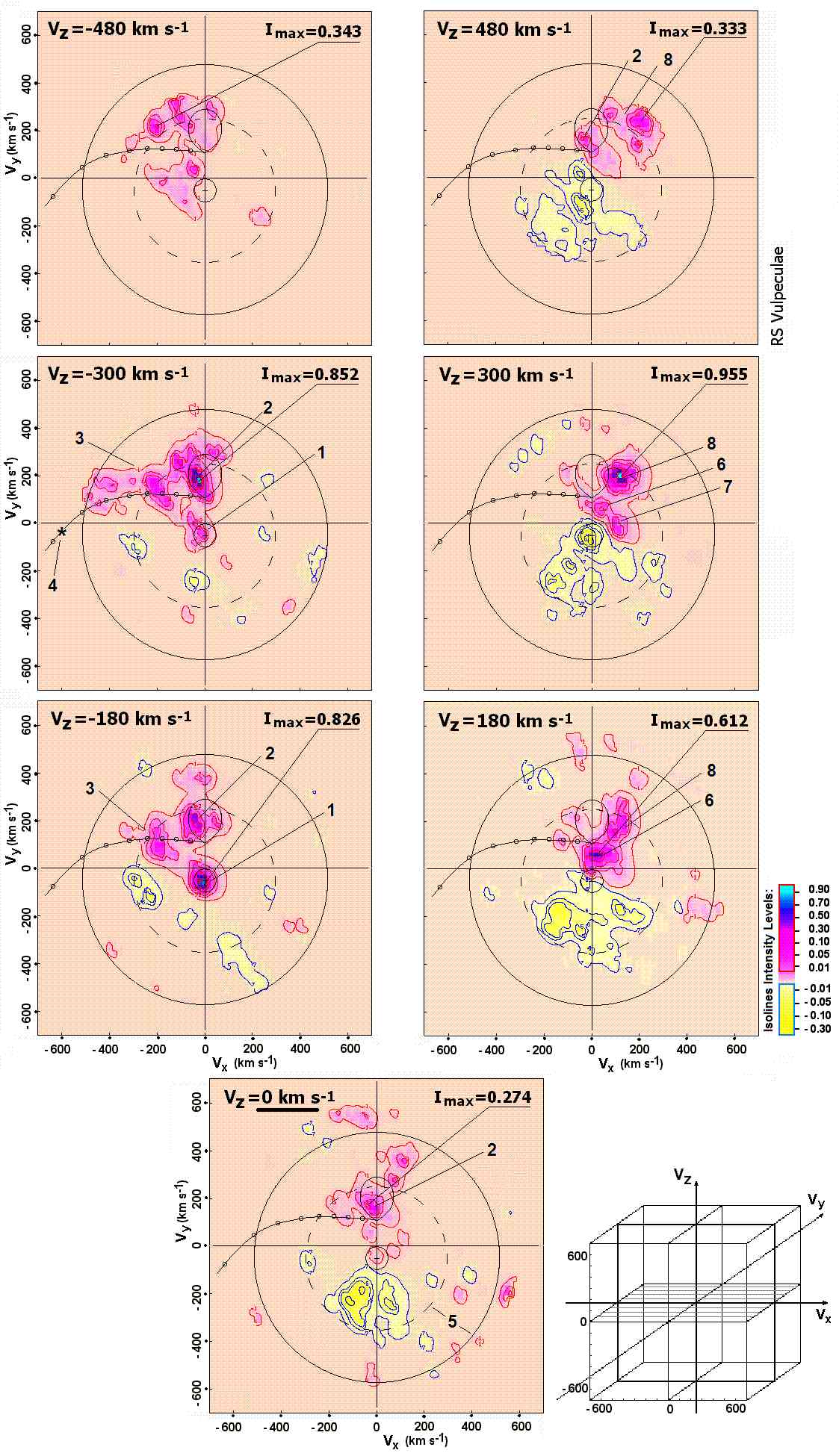}
\caption{Visualization of the seven most interesting ($V_x$,$V_y$) slices in the 3D Doppler Tomogram of RS Vul (1993) displayed symmetrically. The main features are: (1) circumprimary emission, (2) chromospheric emission from the donor star, (3) gas stream, (4) star-stream impact site (the asterisk $\ast$), (5) the predicted locus of accretion disk, (6) localized region between stars, (7) other localized region, and (8) high intensity region with velocity close to that of the donor star.
}
\label{f4}
\end{figure*}

\begin{figure*}
\figurenum{5}
\center
\epsscale{0.7}
\includegraphics[height=200mm]{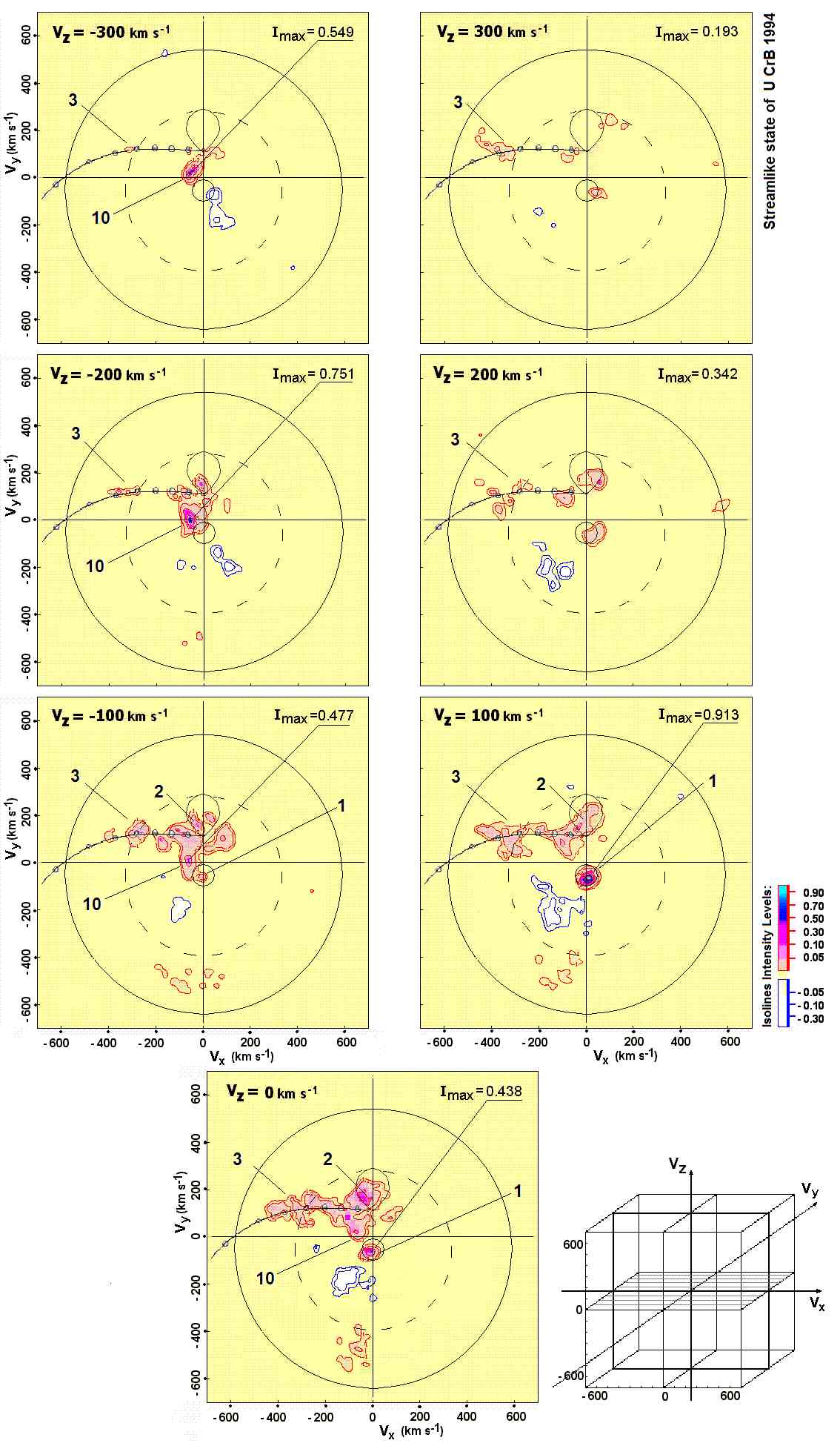}
\caption{Visualization of the seven most interesting ($V_x$,$V_y$) slices in the 3D Doppler Tomogram of U CrB (1994) in the streamlike state displayed symmetrically. The main features are: (1) circumprimary emission, (2) chromospheric emission from the donor star, (3) gas stream, (10) a high-velocity jet in the Vz direction (see \citet{agafonovetal06,agafonovetal09}. 
}
\label{f5}
\end{figure*}

\begin{figure*}
\figurenum{6}
\center
\epsscale{0.9}
\includegraphics[height=200mm]{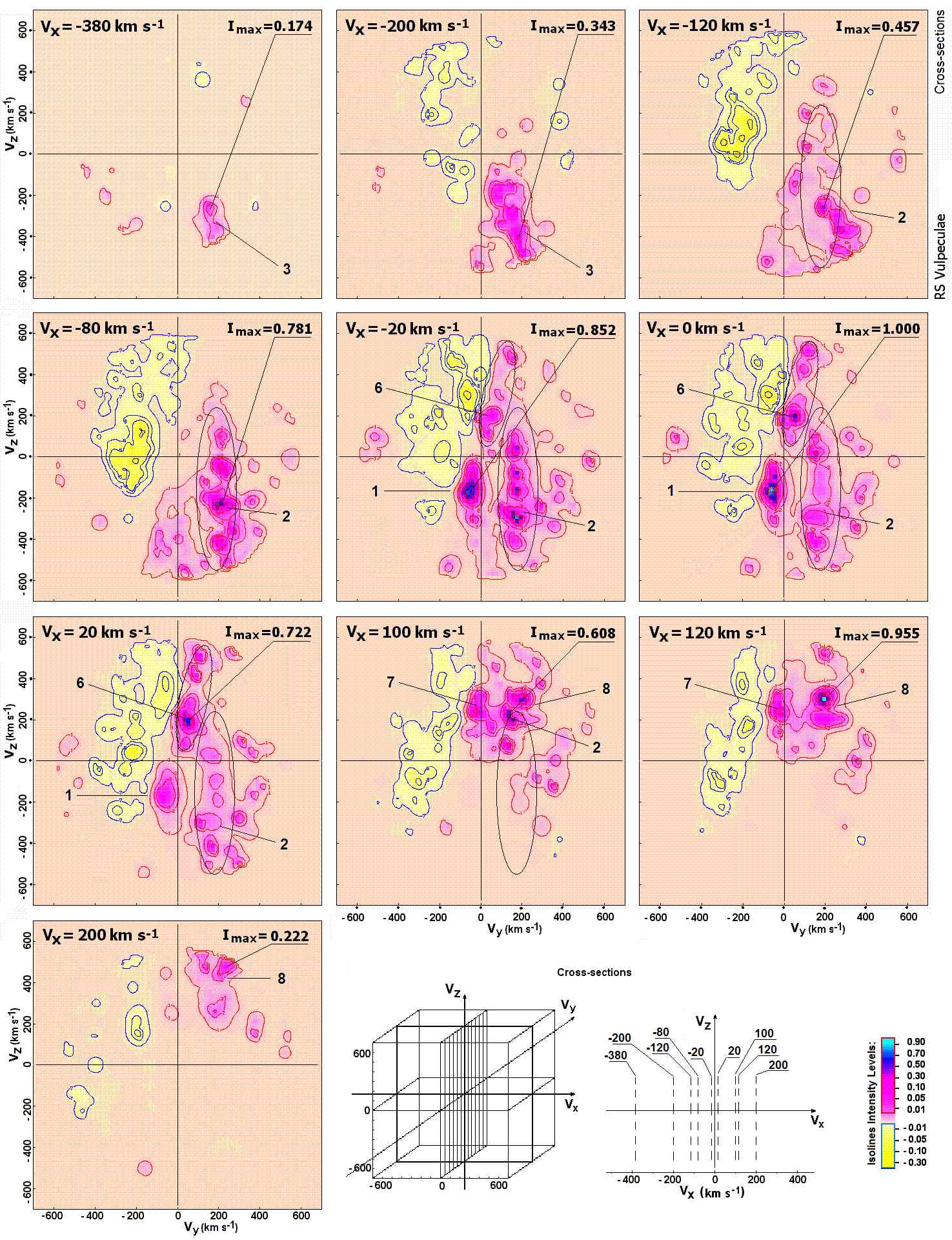}
\caption{Cross sections of the RS Vul 3D Doppler tomogram in the ($V_y$,$V_z$) plane for $V_x$= -380, -200, -120, -80, -20, 0, 20, 100, 120, 200 km~s$^{-1}$  with intensity levels of ± 0.01; 0.05; 0.10; 0.30; 0.50; 0.70; 0.90. The numbered features correspond to those in Figure 3. The maps for $V_x$ values from +200 to +380 km~s$^{-1}$  contain no information.
}
\label{f6}
\end{figure*} 

\begin{figure*}
\figurenum{7}
\center
\epsscale{0.9}
\includegraphics[height=200mm]{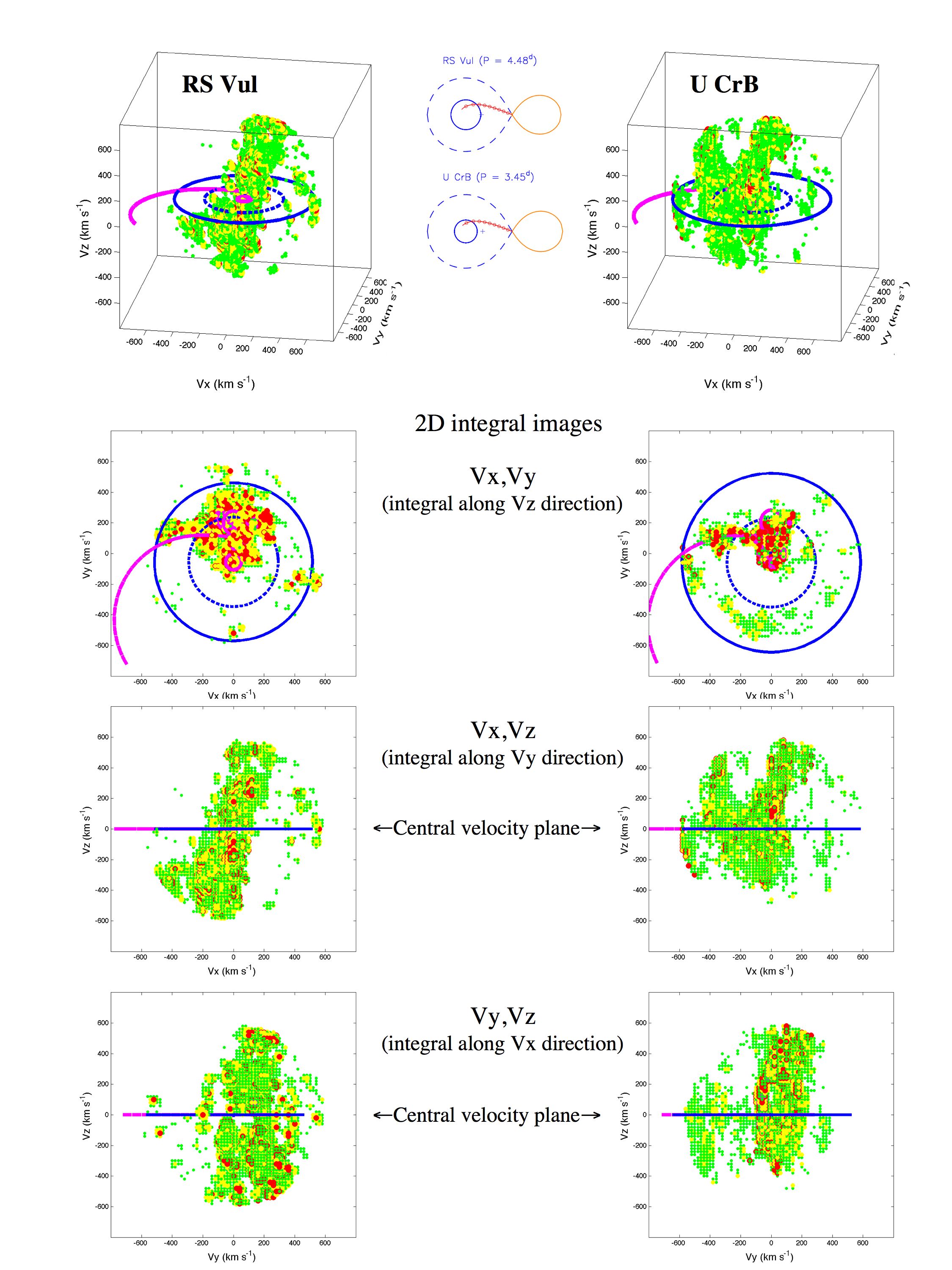}
\caption{The 3D tomograms of RS Vul and U CrB displayed in a 3D velocity cube (top row) together with the 2D integral images for the ($V_x$,$V_y$) plane (second row), ($V_x$,$V_z$) plane (third row), and ($V_y$,$V_z$) plane (last row).  The 2D integral image in the ($V_x$,$V_y$) plane is equivalent to the 2D Doppler tomogram.
}
\label{f7c}
\end{figure*} 
\clearpage

%Table 1. Characteristics and Locations of Prominent Emission Features in the 3D Tomogram.
\begin{deluxetable}{clccc} 
\tablecolumns{5} 
\tablewidth{0pc} 
\tabletypesize{\scriptsize}
\tablecaption{Characteristics and locations of prominent emission features in the 3D tomogram} 
\tablehead{  
\colhead{} & \colhead{ } & \multicolumn{3}{c}{Location: central velocity or velocity range (km~s$^{-1}$)} 
} 
\startdata
{No.} & {Emission Feature} & {$V_x$} & {$V_y$}  & {$V_z$} \\
\hline
1 & Circumprimary emission & 0 (-60 to +60)  & -60 (0 to -120) & -160  (0 to -320) \\
2 & Chromospheric emission & 0 (-120 to +120) & 210 (60 to 360) & -160 (-530 to +350) \\
3 & Gas stream flow & -520 to 0 {\tablenotemark{a}} & 100 to 300 {\tablenotemark{a}} & -300 (-500 to -60)  \\
4 & Star-stream impact region & no source found &  &  \\
5 & Locus of the accretion disk & very weak &   &  \\
6 & Localized Region (LR) -- Part 1 (between stars) & 0 & 0 to 100 & 180 (50 to 300),  450 (360 to 540) \\
7 & Localized Region -- Part 2 & 100 & -80 to 30  & 240 (150 to 350) \\
8 & Jet or flare near donor star & 200 (100 to 300) & $\sim$200 & 360 (180 to 540) \\
\enddata
\tablenotetext{a}{corresponds to the predicted ballistic trajectory of the gas stream.}
%\tablenotetext{b}{corresponds to the predicted location of the feature on the 2D ($V_x$,$V_y$) plane.}
\end{deluxetable}


\begin{thebibliography}{}

\bibitem[Agafonov (2004a)]{agafonov04a}
Agafonov, M. I. 2004a, Astron. Nachr., 325, 259

\bibitem[Agafonov (2004b)]{agafonov04b}
Agafonov, M. I. 2004b, Astron. Nachr., 325, 263

\bibitem[Agafonov, Richards \& Sharova (2006)]{agafonovetal06}
Agafonov, M. I., Richards, M. T. \& Sharova, O. I. 2006, ApJ, 652, 1547

\bibitem[Agafonov \& Sharova (2005a)]{agafonov+sharova05a}
Agafonov, M. I. \& Sharova, O. I. 2005a, Astron. Nachr., 326, 143

\bibitem[Agafonov \& Sharova (2005b)]{agafonov+sharova05b}
Agafonov, M. I. \& Sharova, O. I. 2005b, Radiophysics and Quantum Electronics, 48, No.5, 329

\bibitem[Agafonov, Sharova \& Richards (2009)]{agafonovetal09}
Agafonov, M. I., Sharova, O. I. \& Richards, M. T.  2009, ApJ, 690, 1730

\bibitem[Albright \& Richards (1996)]{albright+richards96}
Albright, G. E. \& Richards, M. T. 1996, ApJ, 459, L99

\bibitem[Bisikalo et al. (2000a)]{bisikaloetal00a}
Bisikalo, D. V., Boyarchuk, A. A., Kuznetsov, O. A., \& Chechetkin, V. M.  2000, Astronomy Reports, 44, 26

\bibitem[Bisikalo et al. (2000b)]{bisikaloetal00b}
Bisikalo, D. V., Harmanec, P., Boyarchuk, A. A., Kuznetsov, O. A., \& Hadrava P, P. 2000, A\&A, 353, 1009

\bibitem[Blondin, Richards \& Malinkowski (1995)]{blondinetal95}
Blondin, J. M., Richards, M. T., \& Malinkowski, M.  1995,  ApJ, 445, 939

\bibitem[Holmgren (1989)]{holmgren89}
Holmgren, D. 1989, Space Sci Rev. 50, 347

\bibitem[Hutchings \& Hill (1971)]{hutchings+hill71}
Hutchings, J. B. \&  Hill, G. 1971, ApJ, 166, 373	

\bibitem[Kaitchuck et al. (1994)]{kaitchucketal94}
Kaitchuck, R. H., Schlegel, E. M.,  Honeycutt, R. K., Horne, K.,
Marsh, T. R., White, J. C., \& Mansperger, C. S. 1994,  ApJS, 93, 519

\bibitem[Kempner \& Richards (1999)]{kempner+richards99}
Kempner, J. C., \& Richards, M. T. 1999, ApJ, 512, 345 

\bibitem[Kuznetsov et al. (2001)]{kuznetsovetal01}
Kuznetsov, O. A., Bisikalo, D. V., Boyarchuk, A. A., Khruzina, T. S., \& Cherepashchuk, A. M. 2001, Astronomy Reports,  45,  872

\bibitem[Lubow \& Shu (1976)]{lubow+shu76}
Lubov, S. H. \& Shu, F. H. 1976, ApJ, 207, L53

\bibitem[Richards (1992)]{richards92}
Richards, M. T. 1992, ApJ, 387, 329

\bibitem[Richards (1993)]{richards93}
Richards, M. T. 1993, ApJS, 86, 255

\bibitem[Richards (2001)]{richards01}
Richards, M. T. 2001, Astrotomography, Indirect Imaging Methods in Observational Astronomy, Edited by H. M. J. Boffin, D. Steeghs and J. Cuypers, Lecture Notes in Physics, 573, 276

\bibitem[Richards (2004)]{richards04}
Richards, M. T. 2004, Astron. Nachr., 325, 229

\bibitem[Richards \& Albright (1996)]{richards+albright96}
Richards, M. T., \& Albright, G. E. 1996, in  Stellar Surface Structure, ed. K. Strassmeier and J. Linsky (Dordrecht: Kluwer), 493

\bibitem[Richards \& Albright (1999)]{richards+albright99}
Richards, M. T. \& Albright, G. E. 1999, ApJS, 123, 537

\bibitem[Richards, Albright \& Bowles (1995)]{richardsetal95}
Richards, M. T., Albright, G. E. \& Bowles, L. M. 1995, ApJ, 438, L103

\bibitem[Richards \& Ratliff (1998)]{richards+ratliff98}
Richards, M. T., \& Ratliff, M. A. 1998, ApJ, 493, 326

\bibitem[Richards et al. (2003)]{richardsetal03}
Richards, M. T., Waltman, E. B., Ghigo, F., \& Richards, D. St. P. 2003, ApJS, 147, 337

\bibitem[Sharova (2006)]{sharova06}
Sharova, O. I. 2006, Proc. of the IAU, {\it Planetary Nebulae in our Galaxy and Beyond}, IAU Symp. 234, 507

\bibitem[Skilling \& Bryan (1984)]{skilling+bryan84}
Skilling, J.  \& Bryan, R. K. 1984, MNRAS, 211, 111

\bibitem[Stern et al. (1992)]{sternetal92} 
Stern, R. A., Uchida, Y., Tsuneta, S., \& Nagase, F. 1992, ApJ, 400, 321

\bibitem[Umana et al (1998)]{umanaetal98}
Umana, G., Trigilio, C., \& Catalano, S. 1998, A\&A, 329, 1010

\bibitem[Vesper \& Honeycutt (1993)]{vesper+honeycutt93}
Vesper, D. N., \& Honeycutt, R. K. 1993, PASP, 105, 731
	
\bibitem[White \& Marshall (1983)]{white+marshall83}
White, N. E., \& Marshall, F. E. 1983, ApJ, 268, L117

\end{thebibliography}
\end{document}